\documentstyle[12pt,aaspp4,epsfig]{article}

\def\ea{et al. }
\def\eg{{\it e.g., \,}}

\def\be{\begin{equation}}
\def\ee{\end{equation}}
\def\cl{\centerline}

\def\bs{\bigskip}
\def\ms{\medskip}
\def\np{\newpage}

\begin{document}
\hfill\today

\baselineskip=14pt
\vspace*{0.5in}

\cl{\Large\bf RXTE View of the Starburst Galaxies M82 and NGC 253}

\bs
\bs
\bs

\cl{\bf Yoel Rephaeli$^{1,2}$}
\bs

\cl{\it and}
\bs

\cl{\bf Duane Gruber$^{1}$}

\bs

\cl{$^1$ Center for Astrophysics and Space Sciences, University of
California, San Diego}

\cl{$^2$ School of Physics and Astronomy, Tel Aviv University, Tel Aviv}
\bs
\bs
\bs
\bs

\begin{abstract}

The two nearby starburst galaxies M82 and NGC 253 were observed for
$\sim 100$ ksec over a 10-month period in 1997. An increase of the M82
flux by a factor $\sim 2$ was measured during the period July-November,
when compared with the flux measured earlier in 1997. The flux measured
in the field centered on M82 includes $\sim 38\%$ of the emission from
the Seyfert 1 galaxy M81. The best-fitting model for the earlier
emission from M82 is thermal with $kT \simeq 6.7 \pm 0.1$ keV. In the high
flux state, the emission additionally includes either an absorbed
second thermal component or absorbed power-law component, with the
former providing a much better fit. A likely origin for the temporal
variability is a single source in M82. The flux of NGC 253, which did
not vary significantly during the period of observations, can be well
fit by either a thermal spectrum with $kT \simeq 3.8 \pm 0.3$ keV, or
by a power law with photon index of $2.7 \pm 0.10$. We have also
attempted fitting the measurements to more realistic composite models
with thermal and power-law components, such as would be expected from
a dominant contribution from binary systems, or Compton scattering of
(far) IR radiation by radio emitting electrons. However, the addition
of any amount of a power-law component, even with cutoff at 20 keV,
only increases chi-square. The 90\% confidence upper limit for power
law emission with (photon) index 1.5 is only 2.4\% of the 2 --- 10 keV
flux of M82; the corresponding limit for NGC 253, with index 2.0, is
48\%.
\end{abstract}
\bs

\keywords{Galaxies: starburst --- galaxies: individual (M82, NGC
253) --- X-rays: galaxies --- radiation mechanisms: non-thermal}
\np

\section{Introduction}

Interest in X-ray emission in starburst galaxies (SBGs) stems from their
diverse radiative activity which is powered by an abundant population of
massive, young stars. The enhanced star formation rate in these gas and
dust rich galaxies leads to intense far-infrared (FIR) radiation from
warm interstellar dust heated by the massive stars. The large number
of X-ray binaries and supernova (SN) remnants, as well as hot winds
driven by SN shocks, imply that X-ray emission from a SBG is more
intense than from a normal spiral (Bookbinder \ea 1980). Moreover,
since SN remnants are prime sites for shock-acceleration of particles
to high energies, Compton scattering of relativistic electrons
by the local radiation fields will enhance X-ray emission in a SBG.
This expectation is heightened by the fact that in a SBG the energy
density of the FIR radiation field can be much higher than that of the
cosmic microwave background radiation (Schaaf \ea 1989, Rephaeli \ea
1991).

The SB phase may occur in various types of galaxies (triggered, perhaps,
by galactic mergers). As in other active galaxies, a SBG may also have
an active, compact nucleus, and one of the primary issues in the study
of SBGs is the possible connection between low-luminosity Active
Galactic Nuclei (AGN) and SB activities. Since both phenomena may occur
in at least some active galaxies, it is important to determine some of
their observational manifestations in order to better characterize
SBGs. This obviously is a prerequisite in the assessment of the
significance of the SB phase in galactic evolution, and its
cosmological ramifications, such as the enrichment of the intergalactic
medium by metal-rich gas ejected during the SB phase, and the
contributions of SBGs to the X-ray background (\eg Rephaeli \ea 1995).

High energy phenomena are triggered by the massive star formation
activity, and because of the obscuration of optical emission and
considerable re-processing of IR emission, X-rays allow a more direct
and penetrating view of the inner regions of SBGs. Detailed X-ray
spatial and spectral information is very valuable for determining the
basic properties of SBGs. However, considerable knowledge on the
emission mechanisms and environments may be gained from good quality
spectral data alone, particularly so if these include temporal
information.

M82 and NGC 253 are the closest SBGs. These `archetypical' SBGs have
been observed with all the major X-ray satellites (e.g., Fabbiano 1988a,
Tsuru \ea 1990, Ohashi \ea 1990, Boller \ea 1992, Matsumoto \& Tsuru
1999, Cappi \ea 1999, Pietsch \ea 2001). Their X-ray emission is quite
substantial, in the range $10^{40} - 10^{41}$ erg/s; reported best-fit
spectra include both thermal (Cappi \ea 1999) and power-law forms (e.g.
Weaver \ea 2000). Both galaxies have already been observed also with
{\it Chandra} (Strickland \ea 2000, Matsumoto \ea 2001, Kaaret \ea
2001). At low X-ray energies, emission from these galaxies extends well
beyond their optical disks, so the SB activity is not manifested only
in the central nuclear SB region ($\leq 1$ kpc). However, even in these
nearby galaxies the characteristics of the emission have not yet been
determined in much detail. It is important to determine the relative
contributions of the SB-powered versus nuclear-powered emission, and
the respective significance of thermal and nonthermal processes.

To better establish the X-ray properties of M82 and NGC 253, we have
initiated observations with the Proportional Counter Array (PCA) and
the High Energy X-ray Timing Experiment (HEXTE) aboard the Rossi X-ray
Timing Explorer (RXTE) satellite. Our main motivation has been to use
the good temporal and wide-energy capabilities of the RXTE in order
to assess the possible role of an AGN-like variable emission, and the
level of contribution of Compton scattering to the integrated emission
in SBGs. These RXTE observations and the results of the spectral and
temporal analysis are reported here.

\section{Instrumentation and Observations}

The RXTE was inserted into low earth orbit on 1995 Dec 31. The PCA was
described by Jahoda \ea (1996) and the HEXTE by Rothschild \ea (1998).
The PCA has energy resolution of 16\% at 6 keV and a useful energy range
of 2.5 keV to about 25 keV, limited at the high end by uncertainties of
determining the internal background. The HEXTE has resolution of 15\% at
60 keV and an energy range from 15 to 250 keV. For the present
observations the errors of background determination with HEXTE were
negligible. For both instruments calibration of energy response has
been extensive and is accurate to better than 2\%. The instrumental
time resolution of microseconds was not needed, but the very flexible
RXTE scheduling of observations was exploited to permit sampling on all
time scales from one orbit (90 minutes) to ten months. Both the PCA and
HEXTE are non-imaging with field-of-view (FOV) one degree.

Over the period February - November 1997, thirty one-orbit observations
each were made of M82 and NGC 253. To sample possible variability
as well as temporal changes through the entire campaign, a somewhat
intricate plan (Table 1) was devised consisting of four observing
sequences. Eight of the orbits were spaced over one day, seven orbits
over one week, another seven over eight weeks, and the remaining eight
over 42 weeks. The spacing of observations in each sequence was not
uniform but moderately irregular: each sequence was divided into 15
equal segments, and observations were placed in either seven or eight
of these segments in patterns based on the cyclic difference set of
size 15 (Baumert 1971). Such patterns allow uniform sensitivity to
variability on a somewhat broader range of time scales or frequencies
than with evenly spaced samples. With the two longest sequences the
corresponding M82 and NGC 253 observations occurred within a day of
each other. Three PCA detectors were employed for all observations.

Analysis of M82 was complicated by variability and by the presence of
another known source with appreciable X-ray emission, M81, in the FOV.
The separation of M81 by $\sim 34'$ from M82 resulted in a transmission
of about 38\% for M81 in the RXTE fields. Although the RXTE field
rotated considerably during the 10-month campaign, the PCA FOV,
nominally a hexagonal pyramid, is sufficiently conical that the area
presented to M81 varied by not more than 10\%. We have also analyzed
contemporaneous archival ASCA and ROSAT data to provide limits to the
M81 contribution.

\begin{table}[h]

\caption{Observation Sequences}

\begin{center}
\begin{tabular}{|l|l|l|l|}

\hline Duration &  Frequency Range ($\mu$Hz) & M82 Start & NGC 253
Start\\
\hline

1 day    &     23.1 - 81.0 & 16 July & 13 July \\

1 week   &      3.31 - 11.6 & 29 March & 11 April \\

8 weeks  &      0.41 - 1.45 & 28 April & 29 April \\

42 weeks &      0.08 - 0.28 & 2 February & 3 February \\

\hline
\end{tabular}
\end{center}
\end{table}

\subsection{Data Reduction and Background Considerations}

The FITS data were reduced with standard FTOOLS into one-orbit light
curves and average spectra. Net counting rates for the three PCA
detectors averaged $21$ s$^{-1}$  and $0.65$ s$^{-1}$ for M82 and
NGC 253, respectively. The NGC 253 counting rate is not very large
compared to the background estimation error of $0.3$ s$^{-1}$ RMS for
the three detectors, using the latest weak-source background model for
the PCA. An initial display of the PCA spectrum of NGC 253 showed a
steep spectrum below 12 keV and above this energy a fairly
flat spectrum with weak lines at 25, 30, and 50 keV. This upper portion
was clearly identifiable with the activation-induced internal
background spectrum at about 2\% of the average background level. This
spurious component was removed with a 1.9\% correction to the
background level. The resulting net spectrum showed no evidence for
a statistically significant flux above 12 keV either with PCA or HEXTE.
The HEXTE internal background was monitored continuously by beam
switching every 16 seconds during the observation and required no
special treatment.

Field-to-field confusion noise from fluctuations of the diffuse X-ray
background (e.g. Shafer 1983) has been calculated (Gruber \ea 1996) to
have an RMS of 8\% of the background flux, and we calculate this level
to be 0.4 c/s for three PCA detectors. This estimated fluctuation level
is comparable to the NGC 253 counting rate. The 2--10 keV spectrum of
the fluctuations, however, has been shown (Butcher \ea 1997) to have
an index of 1.8 $\pm$ 0.02, which is much flatter than what is observed
here for either M82 or NGC 253. From spectral fitting we determined
that the 90\% upper limit to the fraction of the observed 2-10 keV flux
which could come from a background fluctuation is 16\% for NGC 253 and
less than 1\% for M82. Net flux in the FOV of M82 extends up to
$\sim 50$ keV.

\subsection{Temporal Analysis}

We obtained light curves for the M82 and NGC 253 monitoring observations
by fitting each $\sim 3000$ s observation to a Raymond-Smith (R-S) plasma
spectral model, then integrating the model flux to obtain 2-10 keV flux.
The PCA data do not cover the interval 2-2.5 keV, but the 2-10 keV
measure should still be quite reliable, given the small gap and lack
of noticeable spectral variability.

The thirty NGC 253 one-orbit observations had an average counting rate
below 10 keV of 0.65 $\pm$ 0.03 c/s, with reduced chi-square of about
unity, and the RMS of 0.16 c/s is clearly dominated by counting
statistics. No variability is evident, and a 90\% limit to random
(white) variability at the source is 0.145 c/s, or 22\%.

The light curve of M82, Fig. 1, on the other hand, shows clear
evidence for variability on a variety of time scales. Most notably, it
appears to be dominated by a single event which doubles the flux late
in the monitoring period. We indicate in Fig. 1 the maximum and
minimum contributions of M81 to the observed flux, based on occasional
contemporaneous monitoring of M81 with ASCA (Iyomoto \& Makishima 1999)
and ROSAT (Immler \& Wang 2001). The three ASCA and four ROSAT fluxes
shown in Fig. 1 were determined from archival data using the standard
ftools XSELECT and XSPEC. Obviously, these observed levels do not
necessarily sample the full variability of the M81 flux but clearly
they show that a large increase was extremely unlikely. M82, however,
was not monitored with ASCA or ROSAT during the period of RXTE
observations.  Three ASCA observations just before the RXTE monitoring
show both sizeable variability and general consistency
with the RXTE fluxes. A BeppoSAX observation of M82 on December 6-7
1997, about two weeks after the last RXTE observation
segment, shows the flux level of the earlier RXTE observations,
indicating the end of the elevated flux level. If an example of
flaring, this event, with a pattern of fast rise followed by steady
emission then fast decay is unusual for X-ray transients, which
commonly show a fast rise and slow decay. Abrupt state changes of
X-ray emission are seen, however, in the microquasar sources, such
as J1655-40 and 1915-05 (Belloni 1998, and references therein),
which are generally regarded as black hole candidates. Thus a very
interesting possibility is that of an outburst from such a source.
Using earlier ASCA data Matsumoto \& Tsuru (1999) and
Ptak \& Griffiths (1999) have already inferred the existence of a single
strong variable black hole source in M82. The most obvious explanation,
then, is a sudden outburst from a single very luminous source or,
much less likely, coincident outbursts from more than one source.
One should note, however, that the early data before this
event had variability of 10\% RMS above counting statistics.

In addition to the single-outburst scenario, we analyzed whether the
observed power density spectrum resembles that of known variable
sources. In particular, we tested the observed power density for a
power-law decline of spectral density with increasing frequency, i.e.
a ``red'' spectrum such as seen in the stellar-size black hole Cyg X-1
(Nolan \ea 1981), in several AGNs (Markowitz \& Edelson 2001), and in
HMXRBs generally (Belloni \& Hasinger 1990).

In treating the observed M82 variability as a red noise process, the
measured flux was assumed to be a stochastic variable with power over a
broad range of frequencies, increasing with decreasing frequency.
Although a fine-grained frequency analysis was not possible because of
the modest number of samples and marginal statistics, the design of the
observing schedule nevertheless permitted a particularly simple
analysis: for each of the four sets of observations (specified in the
beginning of this section) an average flux for the seven or eight
samples was calculated. Values of $\chi^2$ were then computed based
solely on counting statistics, which resulted in formal values that were
unacceptably large, indicating extra variance. Then it was assumed that
for each set the flux was intrinsically variable with stationary mean
and some variance, or -- more conveniently -- RMS.  We adjusted this
assumed RMS and determined the values at which the $\chi^2$ probability
assumed the values 0.05, 0.155, 0.5, 0.745 and 0.95, thus determining
the best fit value for the RMS and the 68\% and 90\% confidence bounds.
The best fit value and standard error for the RMS were normalized to the
average flux level to obtain a fractional measure of variability.
Finally, an average frequency and frequency range were assigned to
each variability estimate, thus producing the values in Table 2 and
Fig. 2.

\begin{table}[t]

\caption{Broad-Band M82 Variability vs. Frequency}

\begin{center}

\begin{tabular}{|c|c|c|c|}

\hline Frequency ($\mu$Hz)  &  RMS   &  $\sigma$ &  90\% Interval \\
\hline

  0.15    &    0.464  &  0.113     &  [0.303, 0.843] \\

  1.10    &    0.167  &  0.046     &  [0.101, 0.328] \\

  7.5     &    0.040  &  0.022     &  [0.000, 0.085] \\

  55.     &    0.032  &  0.013     &  [0.010, 0.071] \\

\hline
\end{tabular}
\end{center}
\end{table}

\begin{figure}[h]
\centerline{\psfig{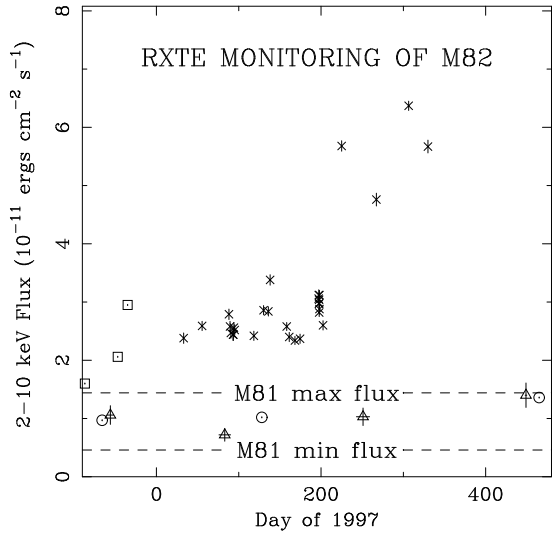}}
\figcaption{Observed fluxes of M82 and M81 in 1996, 1997 and 1998. RXTE
measurements include M81, which was in the RXTE field at 38\%
transmission. The 30 measurements from the RXTE monitoring campaign
are marked with x's, ASCA fluxes of M82 and M81 are marked with
circles and squares, respectively, and the ROSAT fluxes of M81 are
marked with triangles.  The observed ROSAT and ASCA fluxes of M81 have
been reduced by the factor 0.38 to correct for the RTE transmission.
Reasonable limits to the M81 contribution, as used in the spectral
analysis, are shown by dashed lines.}
\end{figure}

The RMS clearly decreases strongly with frequency. Fitting to a
power-law dependence we obtain an excellent fit with a correlation
coefficient of -0.98 and an index of $0.50 \pm 0.02$. This corresponds
to an index of unity for the power spectrum, or what is sometimes called
a flicker noise spectrum. These observations, then, are consistent with
a mildly red noise process extending without flattening to the lowest
frequency of about $0.1 \mu$Hz. Such a determination contradicts the
initial assumption of a stationary process, but not strongly (see
Deeter 1984).

\begin{figure}[t]
\centerline{\psfig{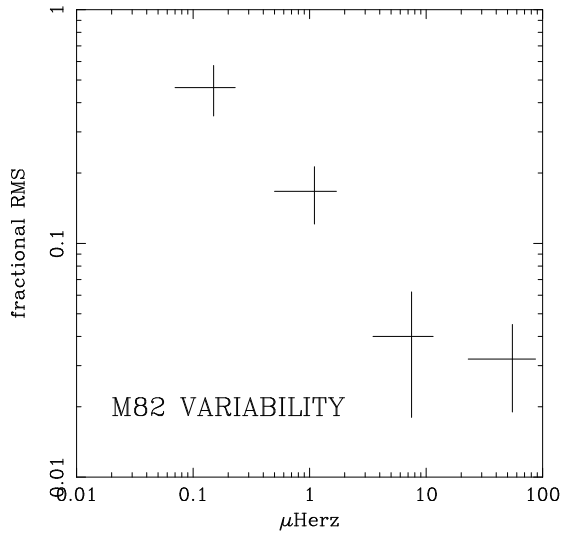}}
\figcaption{Power spectrum of flux variability in M82. This spectrum
may indicate a flicker noise process at periods longer than a day.}
\end{figure}

Alternatively analyzing the M82 light curve as having two levels, early
(first 26 samples) and late (last 4), we assume that the higher emission
at late times represents an addition to the average flux observed at
earlier times. We first examine the early emission. Versus time,
there is a fitted rise by 20\% from February to July which is formally
significant to 3$\sigma$, but has a correlation coefficient of only
0.52. Therefore we consider the trend to be an incidental effect of
more rapid source variability. Assuming a stationary process, we
determine a fractional RMS of $0.106 \pm 0.014$, or a 90\% confidence
region [0.083, 0.130]. We estimate the limit to the transition time
between the two levels of flux to be about 25 days.

Standard power spectral analysis was used to test whether the observed
RXTE 2--10 keV fluxes were more consistent with a step function change
in time or with the flicker noise spectrum suggested by Fig. 1. Since
the sampling was very uneven, this was done by constructing a comparison
time series in which the low and high state data points were each
replaced with their respective averages. Power spectra were calculated
and the resulting power estimates averaged over four broad bands
approximating those in Fig. 1 in order to suppress sampling noise.
Within errors of about 25\% per average, the two processes, step change
and flicker noise, cannot be distinguished.

We then tried to determine if the four late samples have a pattern of
variability distinct from the 26 earlier samples. We find for the last
four samples that the fractional RMS is 0.098, about the same as for the
early data. Analyzing versus frequency as above, we compute a best-fit
index of 0.29, which is, however, not significantly different from zero
(P=0.19) or from 0.5 (P=0.30). From temporal analysis alone it
cannot be determined if the noise process governing the late emission
is the same as, or rather different from that of the earlier emission.

In summary, from analysis of the noise properties, we find that
the M82 variability determined from the entire sample is broadly
consistent with a power density spectrum of power law form with index 1.
An effort to determine if the late high emission had distinct temporal
properties, which could indicate a distinct origin, failed because of
the small sample of only four observations.

\subsection{Spectral Analysis}

Spectral analysis was done with the use of XSPEC and additional
statistical tools. Latest versions of the PCA and HEXTE response
matrices were used, and a $\sim 1\%$ systematic error was assumed in
order to allow for errors in the calibration of the PCA response matrix.
PCA data above 25 keV were excluded to avoid errors in background
modeling. Nevertheless, 3\% negative residuals at 4-5 keV, which is
the peak of the background spectrum, were attributed to background
estimation errors and modeled specially in the fitting. PCA data below
2.5 keV were ignored due to uncertainty of the response matrix in the
first few channels. These data are thus insensitive to the 0.7 keV
thermal component reported by Cappi \ea (1999), Weaver, Heckman \&
Dahlem (2000), and Dahlem \ea (2000), and we ignore this component.
HEXTE data above 50 keV were felt to add only noise in the fits and
were not used for the results reported here.
\ms

\cl{\bf NGC 253}

NGC 253, observed at an average flux level of 2.4$\times$ 10$^{-12}$
erg cm$^{-2}$ s$^{-1}$ in the $2 - 10$ keV band (or $\sim 0.1$
milliCrab), was significantly detected only by the PCA. Thus HEXTE
data were not used for spectral fitting. Since no variability could be
determined for this source, data from all thirty $\sim$3000 s samples
were coadded, as well
as the background estimates, to produce a mean spectrum. Net detection
was at about 20$\sigma$ significance, which was strong enough
for simple spectral modeling. We tried power-law and thermal models. For
the thermal case we tested the PCA data below 25 keV with both an R-S
thermal model and thermal bremsstrahlung with iron line. Both thermal
models yielded a temperature of 3.8 $\pm$ 0.3 keV. The best-fit Fe
abundance in solar units, $Z_{\rm Fe}$, with R-S was 0.16 $\pm$ 0.08,
and with bremsstrahlung the equivalent width was 160 $\pm$ 130 eV. In
neither thermal case was an absorbing column required by the data.
Values of $\chi^2$ are close to 34 for 50 degrees of freedom; although
the reduced $\chi^2$ is low, it is not sufficiently low to indicate a
serious overestimate of errors. An unabsorbed power law model gave a
best-fit number index of 2.7 $\pm$ 0.1, and iron equivalent width of 340
$\pm$300 eV. (The R-S fit, which tests for several iron lines plus
an edge, is more constraining than a fit for a single line.) Although
the power law was formally a good fit, with $\chi^2 \simeq 41.2$
for 50 degrees of freedom, the residuals show correlated departures
which indicate that the acceptable $\chi^2$ is an artifact of the
oversampling in energy. No such correlations are seen in the residuals
to the thermal fits. Relative likelihoods from Poisson statistics were
calculated from the data and fits, and either thermal form is about
1000 times more likely than the unabsorbed power law. An absorbing
column of $N_{\rm H} = 2.8\times 10^{22}$ cm$^{-2}$, however, brings the
goodness-of-fit close to the thermal fits. In this case the best-fit
index steepens to 3.1 $\pm$ 0.2. The fitting was performed on the
interval 2 -- 22 keV, which includes many channels above 12 keV, where
the source is no longer visible. However, the best fit parameters and
their errors are very insensitive to a variation of the upper cutoff
energy, even when it is lowered to 10 keV.

Because of the faintness of NGC 253, the best-fit spectral parameters
show a moderate dependence on the background subtraction. The value of
$\chi^2$ changes significantly when the background level is adjusted
by $\pm$0.5\%, for which we obtain corresponding changes for best-fit
temperature of 0.7 keV, $Z_{\rm Fe} \simeq 0.05$ for abundance, and 0.15
for spectral index.
These systematic errors resulting from the uncertainty of background
subtraction are less than twice the formal statistical errors from
counting statistics. Somewhat smaller errors can be associated with a
fluctuation of the X-ray background. At the 90\% upper limit of 16\%
of the net 2--10 keV flux attributable to such a fluctuation, the
best-fit temperature is then smaller by 0.4 keV and $Z_{\rm Fe}$ is
larger by 0.03.
\ms

\cl{\bf M82}

Spectral analysis of M82 includes consideration of the maximum and
minimum possible contribution from the confusing source M81. From
these limiting cases we show that the spectral analysis is not
seriously compromised by M81. It was particularly interesting to
observe the difference between the late high-flux emission and the
earlier low emission (Fig. 1), with no allowance being made yet for
any contribution from M81. This lower flux averaged
2.9$\times$10$^{-11}$ erg cm$^{-2}$ s$^{-1}$ in
the 2-10 keV band, or $\sim 1.3$ milliCrab, and the average for the four
high samples was almost exactly double this level. We analyzed spectra
summed over the whole campaign, the 26 (low) earlier observations, the
last 4 (high) observations, and finally obtained a difference spectrum.
The early, late and difference spectra all show an overwhelming
preference for an exponential shape over an unabsorbed power law --
the reduced $\chi^2$ using the power law form lies in the range 3--4.
Even a small power-law component results quickly in significantly
increased $\chi^2$: a 15\% contribution increases $\chi^2$ by 12 in
all but the low-flux spectra. A comparison of Tables 3, 4, \& 5 below
shows the rapid worsening of $\chi^2$ with increasing power law
fraction as represented by the power-law M81 contribution.

Results for the best-fit thermal unabsorbed and absorbed models for the
{\it total} emission in the FOV of M82, i.e. without allowance for an M81
contribution, are presented in Table 3. To represent thermal emission we
employed both a Raymond-Smith plasma model and a thin thermal
bremsstrahlung with a separate single iron line. Empirically, we find
little difference between the two forms: values of $\chi^2$ were almost
identical, as were the best-fit values of kT, and the iron line
centroid floated to values near 6.7 keV. In the `quiescent'
low-brightness phase, $kT = 7.35 \pm 0.07$ keV, and the Fe line
strength is represented by an equivalent width of 120 $\pm$ 30 eV, or
with the R-S model by of $Z_{\rm Fe} = 0.10 \pm 0.01$. The ``outburst'',
or difference emission (high-low) is best-fit by an absorbed thermal
spectrum with $kT = 4.72\pm 0.17$ keV, but has a very weak iron line of
equivalent width 50 $\pm$ 70 eV, or in the R-S model $Z_{\rm Fe}=
0.02 \pm 0.02$.

\begin{table}[h]

\caption{Summary of Thermal Spectral Fits to the Full Emission in the
FOV of M82}

\begin{center}
\begin{tabular}{|l|c|c|c|c|c|c|}

\hline Dataset & $\chi^2$ & $N_{\rm H}$ cm$^{-2\/a}$ & kT$^a$ & Abundance 
$^a$&
Fe Eq. Width & Flux$^b$ \\
  & (110 df) &(10$^{22}$ cm$^{-2}$) &(keV)&(solar)& (eV) & \\

\hline

All    & 139 & 0 & $7.08 \pm 0.08$ & $0.105\pm 0.007$ & 130 $\pm$ 20 & 3.36 
\\
Low    & 113 & 0 & $7.35 \pm 0.07$ & $0.102\pm 0.009$ & 120 $\pm$ 30 & 2.95 
\\
High   & 140 & 0 & $6.59 \pm 0.09$ & $0.103\pm 0.012$ & 150 $\pm$ 30 & 5.80 
\\
High-Low & 140 & 0 & $5.91 \pm 0.16$ & $0.097\pm 0.023$ & 170 $\pm$ 70 & 
3.36 \\
High-Low & 98 & $2.7 \pm 0.4$ & $4.72 \pm 0.17$ & $0.020\pm 0.020$ & 50 
$\pm$
70 & 3.93\\
\hline
\end{tabular}
\end{center}
\tablenotetext{a}{All quoted errors are at the 68\% confidence level.}

\tablenotetext{b}{Flux in 2-10 keV band, in units of $10^{-11}$ erg
cm$^{-2}$ s$^{-1}$}
\end{table}
The above analysis was then repeated assuming maximal contribution from
M81, given its variability. The long term 2-10 keV flux of M81 (as
measured by six X-ray satellites) seems to have varied by a factor of
$\sim 3$ during the last twenty years (see Fig. 2 in Pellegrini \ea
2000). In 1997, when the RXTE observations were made, ASCA
measured a flux which varied in the range $\sim (2.4-3.4)\times
10^{-11}$ erg cm$^{-2}$ s$^{-1}$ in this energy band (Iyomoto \&
Makishima 1999). This is also the largest variation observed over a
single year. A slightly higher flux, $3.8\times 10^{-11}$ erg cm$^{-2}$
s$^{-1}$, was measured by BeppoSAX in June 1998 (Pellegrini \ea 2000);
their best-fit continuum spectrum was determined to be a power-law with
photon index $\alpha = 1.86$. The flux range measured by ASCA implies
-- with the 38\% transmission -- that the M81 emission constitutes
$\sim 31\%-44\%$ of the counting rate in the low samples, and $\sim
16\%-22\%$ of the high average. The results of repeating the fits of
Table 3 in the presence of the highest ASCA-measured flux from M81,
assuming a power-law spectrum with index 1.8, are collected
in Table 4. A comparison of Tables 3 and 4 indicates that the best-fit
spectral parameters do not depend strongly on the uncertain contribution
of M81. The best-fit thermal R-S model (Fig. 3) to the low-state
emission has kT reduced by 10\% from 7.3 keV (without explicit
accounting for emission from M81) to 6.7 (with maximal M81 flux), and
has a somewhat higher value for Fe abundance, $Z_{\rm Fe} = 0.14 \pm
0.01$. This value of the temperature is appreciably lower than the
best-fit value deduced from the BeppoSAX measurements of Cappi \ea
(1999), $kT = 8.20 \pm 0.59$ keV; their value for the $Z_{\rm Fe}$,
$0.07 \pm 0.02$, is very low and formally inconsistent with these RXTE
results.

We tried also a composite thermal plus power law spectrum to simulate an
average Galactic binary X-ray spectrum (White et al. 1983), which
has recently been used to model the contribution of binaries to X-ray
emission in SBGs (Persic \& Rephaeli 2002). A full representation of
such a model requires specifying various components with a range of
parameter values (including cutoff energies), most of which are poorly
known. Clearly, only simple spectral modeling is warranted with our
RXTE measurements; indeed, even with the more constraining
spatially-resolved spectral analysis afforded by the
joint ROSAT and ASCA datasets, only a simple combination of thermal plus
power-law components was used to characterize the integrated emission of
massive binary systems in SBGs (Dahlem \ea 1998, Dahlem \ea 2000,
Weaver \ea 2000). Adding a cut-off power-law spectrum with a photon
index of 1.5 (Dahlem \ea 1998) we found that any amount
of such emission increased $\chi^2$ from the pure thermal case, and we
obtained a 90\% confidence limit of 2.4\% of 2--10 keV flux in this
component. This result is not very sensitive to the assumed level of
M81 emission.

We consider also absorbed thermal and power law spectral forms for the
extra emission in the late four samples (Tables 3, 4 and 5). The only
acceptable $\chi^2$ was obtained with an absorbed thermal model (Fig.
4) with kT = 3.80 $\pm$ 0.15 keV, $Z_{\rm Fe} = 0.04$, and column
$N_{\rm H}=4.3\times 10^{22}$ cm$^{-2}$. The best-fit power law to the
residual emission was steep, index greater than 3. Qualitatively
similar results were obtained whether the fits were performed after
subtraction of the lower M81 ASCA flux, or the higher BeppoSAX fluxes.
As can be seen in the middle panel of Fig. 4, residuals to a fit
which neglects absorption leave a pattern of correlated residuals over
the entire span of 3 -- 20 keV.

\begin{figure}[t]
\centerline{\psfig{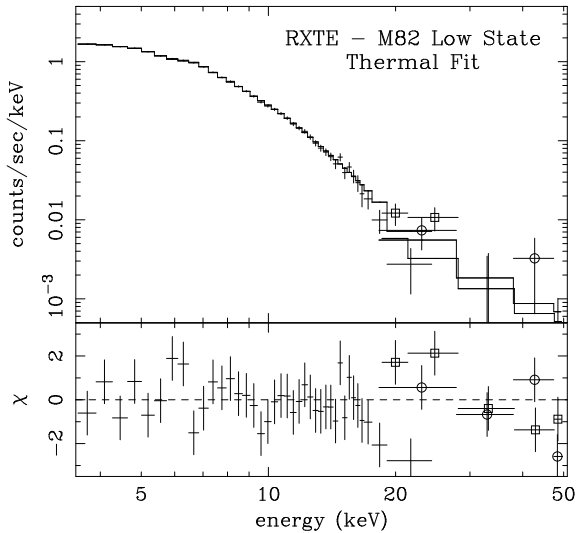}}
\figcaption{Low-state M82 emission and best-fit R-S model with $6.66 \pm
0.08$ keV. HEXTE data are marked with circles (cluster A) and squares
(cluster B).  The M81 contribution is set at the maximum of Fig. 1.
The lower panel shows acceptable residuals to the fit.}
\end{figure}

\begin{table}[h]

\caption{Results of Thermal Spectral Fits for M82, with High M81
Component}

\begin{center}

\begin{tabular}{|l|c|c|c|c|c|c|}

\hline Dataset & $\chi^2$ & $N_{\rm H}$ & kT & Abundance & Fe Eq. Width &
Flux$^a$\\
& (110 df) & (10$^{22}$ cm$^{-2})$ & (keV) &(solar)& (keV) & \\
\hline

3.40 \\
All    & 175 & 0 & $6.46 \pm 0.06$ & $0.138\pm 0.008 $ & 180 $\pm$ 30 & 1.92 
\\
Low    & 137 & 0 & $6.66 \pm 0.08$ & $0.141\pm 0.010$ & 190 $\pm$ 60 & 1.55 
\\
High   & 153 & 0 & $6.22 \pm 0.09$ & $0.120\pm 0.012$ & 200 $\pm$ 40 & 4.40 
\\
High-Low & 159 & 0 & $5.17 \pm 0.16$ & $0.147\pm 0.028$ & 220 $\pm$ 80 & 
2.89 \\
High-Low & 100 & $4.3 \pm 0.5$ & $3.80 \pm 0.15$ & $0.037\pm 0.022$ & 110 
$\pm$ 120 & 3.76\\

\hline
\end{tabular}
\end{center}
\tablenotetext{}{See footnotes to Table 3}
\end{table}

\begin{figure}[t]
\centerline{\psfig{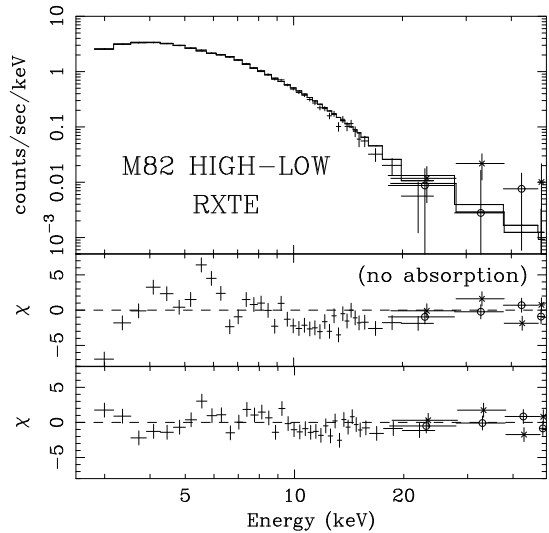}}
\figcaption{The extra emission in the high flux state of M82 and
best-fit absorbed R-S model with $kT = 3.80\pm 0.15$ keV. HEXTE data
are marked as in Fig. 3. The lowest panel shows the acceptable
residuals to the best-fitting absorbed extra emission, while the middle
panel shows a broad pattern of correlated and unacceptable residuals
to an unabsorbed extra component.}
\end{figure}

\begin{table}[h]

\caption{Power Law Spectral Fits to the High State Increment}

\begin{center}
\begin{tabular}{|l|c|c|c|c|c|}

\hline M81 Emission & $\chi^2$ & $N_{\rm H}$ & Index & Fe Eq. Width&
Flux$^a$\\
& (110 df) & 10$^{22}$ cm$^{-2}$& & (eV) & \\
\hline

Not included    & 292 &             0 & $2.38 \pm 0.02$ & 310 $\pm$ 180  & 
2.89 \\
Not included    & 123 & $6.8 \pm 0.6$ & $3.02 \pm 0.06$ & 130 $\pm$ 30 & 
5.23  \\
Included        & 307 &             0 & $2.45 \pm 0.03$ & 390 $\pm$ 90  & 
2.38 \\
Included        & 133 & $8.6 \pm 0.8$ & $3.32 \pm 0.08$ & 170 $\pm$ 90  & 
5.19 \\
\hline
\end{tabular}
\end{center}

\tablenotetext{a}{{\bf Intrinsic, unabsorbed} 2-10 keV power law flux in
units of 10$^{-11}$ erg cm$^{-2}$ s$^{-1}$}
\end{table}

The best-fit thermal forms for the late extra emission have temperatures
that are modestly but significantly lower than for the earlier emission.
If the entire data set samples a single source and process, then this
process must generate somewhat softer spectra at higher emission levels.
We note that whether the extra emission is assumed to have a thermal or
a power law form, the addition of an absorbing column results in a
significant improvement in $\chi^2$, dramatically so for the power law
form. The appearance of an extra column with increased flux levels may
not be unphysical even if the high and low levels result from a single
process, but it is simpler to interpret the need for absorption as an
indication of an obscured source. Thus, based on this spectral analysis
it would appear that the origin of the extra late emission is in a
source, most likely a single source, which did not emit the earlier
lower flux.

A similar high-low analysis has been performed by Matsumoto \& Tsuru
(1999) and by Ptak \& Griffiths (1999), both using essentially the same
ASCA data collected from 1993 to 1996. Matsumoto \& Tsuru give the more
useful summary of spectral fitting of the high-low flux: a thermal form
with best-fit $kT = 11.7 \pm 2.5$ keV, fits slightly better than a
power law with best-fit number index 1.8.  Both forms require strong
absorption with column near $1.4 \times 10^{22}$ cm$^{-2}$. These RXTE
results, with much better high energy sensitivity, strongly favor a
thermal form over power law, but the best fit values for column and
kT, $2.7\times 10^{22}$ and 4.7 keV, respectively, each differ from
the corresponding ASCA value by a factor of two.  Differences of this
scale probably cannot be dismissed as intercalibration errors: ASCA
and RXTE saw something rather different. If the same single source was
seen, it varies on the scale of roughly a year in both column and
general spectral shape.  Much of the absorbing column must lie near to
the source.

\section{Discussion}

Given the varied nature of X-ray emitting environments in SBGs,
significant thermal emission is clearly expected from the relatively
high abundance of various stellar sources such as X-ray binaries,
hot gas in SNRs, and SN-driven winds in the IS space and inner halo.
The main issues that can be addressed based directly on our spectral
analysis have been briefly discussed in the previous section. Here
we elaborate further on the implications of these results.

A detailed comparison between the results of the RXTE and BeppoSAX
observation is complicated, given that the PCA has a larger FOV than that
of the MECS experiment. Also, no temporal comparison is possible because
the observations were made over non-overlapping periods, and due to the
fact that BeppoSAX observed M82 over only two days. It is nonetheless
of considerable interest that BeppoSAX measurements do show some flux
variability in the central region of M82. This is a clear indication that
at least part of the variability seen in the PCA data is intrinsic to
M82, and is not fully attributable to the variable (Seyfert 1) galaxy
M81, which is in the FOV of the PCA but not in that of the MECS. Some,
seemingly periodic, variability of the M81 emission on a timescale of
up to $\sim 2$ days, has been seen recently by BeppoSAX (Pellegrini
\ea 2000). If this is characteristic of M81, then it is clear that the
stronger variability detected by the RXTE over a longer timescale has
a different origin. We also note -- from this general comparative point
of view -- that there is no indication of a second, low-temperature
component in the RXTE data, whose lower threshold energy at $\sim 3.5$
keV makes RXTE insensitive to it.

An important result of our analysis is the large secular variation in
the flux of M82 during the last four observation segments in July 1997.
Such a large temporal change of galactic-scale emission is commonly
attributed to the characteristics of the accretion process onto
a massive black hole. These include also power law noise spectra which
have been observed for several AGN sources (\eg Edelson \& Nandra 1999)
with spectral indices of 1.5 -- 1.7, only mildly inconsistent with the
present case of 1.0 $\pm$ 0.4. The variable emission would then be
expected to have an AGN-like power law energy spectrum with very
significant photoelectric absorption. Indeed, our spectral analysis
yields evidence for the presence of such a component in the variable
emission from M82. The deduced values of $N_{\rm H}$ are rather high,
$\sim (3-20)\times 10^{22}$ cm$^{-2}$ (about an order of magnitude
higher than column densities in Seyfert I galaxies such as M81), perhaps
indicative of a Seyfert II nucleus (\eg Risaliti \ea 1999). However,
the best-fit power law index, $\alpha \sim 3.9$, is
much higher than the more typical value of $\sim 1.8$. The thermal
fit of the residual emission with the high absorption yields a very
low iron abundance, perhaps too low to be acceptable.

Alternatively, the flux increase in July 1997 may possibly be due to
variable emission from an X-ray binary system. If in M82, the high
level of emission ($L \geq 4\times 10^{40}$ erg-s$^{-1}$ in the 2-10
keV band) would make this unusually luminous when compared with the
more common luminosity range ($10^{36}-10^{38}$ erg-s$^{-1}$) of
binaries. Although rare, a few such X-ray luminous binaries are
known (\eg in the spiral galaxies NGC3628 and NGC4631 - Fabbiano
1988b). The spectral shapes of these sources are usually fit with
power-law indices $<2$, or $kT$ values of a few keV. Arguing against
a binary origin for the extra emission is the very low upper limit
of 15 eV for the equivalent width of an Fe K$\alpha$ line. The 320
eV equivalent width of Hercules X-1 Fe K$\alpha$ line (Gruber \ea
2001) reasonably typical of Galactic accreting binaries, is a factor
$\sim 20$ higher.

>From ASCA monitoring in 1993 and 1999 it was concluded (Matsumoto \&
Tsuru 1999, Ptak \& Griffiths 1999) that the M82 flux varied by a
factor $\sim 4$ over this period. They argue that the varying emission
originates from within a central 10'' region, and that if due to
accretion onto a massive black hole, the implied mass is $\geq 460 \,
M_{\odot}$. A compact source in the nucleus with the deduced level of
luminosity, $\sim 4.9\times 10^{40}$ erg/s in the 2-10 keV
band (based on a distance of 3.6 Mpc), would imply that M82 is a low
luminosity AGN. However, we consider the identification of the source of
flux variability with an AGN in M82 to be insecure. Continued
monitoring of the central emission in M82 is needed in order to better
establish its temporal, spectral, and spatial properties. More
recently, high-resolution observations of the central 1' region of M82
with the {\it Chandra} satellite revealed 4 sources whose flux exhibits
significant temporal variability (Matsumoto \ea 2001, Kaaret \ea 2001).
In particular, the 0.5--10 keV flux of one of these sources (CXOM82
J095550.2+694047) varied by a factor of $\sim 7.2$ (during the period
October 1999 -- January 2000). Matsumoto \ea (2001) suggest that this
source is the origin of the variability detected by ASCA. In its high
flux state, the emission from this source is comparable to the high-low
flux we deduce from the RXTE measurements.

Irrespective of the origin of the variable spectral component in M82, it
is important to note that about half of the measured emission is at most
weakly varying. Even though the superposed emission from many variable
sources is non-varying, given the thermal (unabsorbed) character of the
non-varying emission, and previous evidence for its spatial extent, it
is reasonable to conclude that this emission is powered by starburst
activity.

Cappi \ea (1999) have suggested that the main high-temperature
spectral components, which account for most of the observed 2-10 keV
emission in both NGC 253 and M82, are largely due to emission from hot
galactic winds. While we do expect the intense star formation activity
to drive hot galactic winds, it is clear that thermal emission in SBGs
is a superposition of emissions from a population of X-ray binaries,
SN remnants, and galactic winds. The determination of the exact origin
of the main high-temperature spectral component will be possible only
when more sensitive spectral and imaging measurements are made of NGC
253 and M82.

Both BeppoSAX and RXTE find no evidence for variability in the emission
from NGC 253, so this galaxy is not dominated by an AGN. The best-fit
temperature derived from the BeppoSAX measurements (Cappi \ea 1999),
$\simeq 5.75 \pm 0.52$ keV (90\% confidence), is roughly consistent with
our best-fit value of $4.2 \pm 0.4$ (68\% confidence), when the
uncertainty in the background subtraction is included. Even though the
value of the iron abundance we determine here is lower than that
deduced from the BeppoSAX measurements, both are substantially
uncertain and (therefore) in rough agreement.

Power law and thermal models were previously found to provide
acceptable fits to the emission observed by ASCA from both M82 \& NGC
253 (Ptak \ea 1997.) In fact, Moran \& Lehnert (1997), who analyzed 1993
ASCA observations, concluded that the main component of the emission in
M82 is nonthermal, and based on the similar values of the X-ray and
radio power law indices, they suggested that the emission is from
Compton scattering of relativistic electrons by the local radiation
fields. Indeed, the low iron abundances we deduce here are puzzling
given the intense star formation activity, and the associated
processing of metal enriched interstellar gas in these galaxies.
Higher abundances are obtained not only when the dominant spectral
component of the continuum emission is a power law, but in any case
where nonthermal emission contributes appreciably to the overall
emission. For example, if the 2-10 keV flux in the power law component
is taken to be 25\% of the thermal component, then the iron abundance
in M82 increases from its best-fit value $0.14$ (Table 4) to 0.23.

It has previously been shown by Goldshmidt \& Rephaeli (1995) that
Compton scattering of radio producing relativistic electrons by the
far infrared radiation field can account for the substantial high
energy ($50 - 150$ keV) emission detected by OSSE (aboard CGRO) from
NGC 253 (Bhattacharya \ea 1994). The similarities in radio properties
of M82 and NGC 253 and their common starburst nature are sufficiently
strong indications that Compton scattering may play an appreciable
role also in M82. However, neither the BeppoSAX PDS nor the HEXTE
observations were sufficiently sensitive to clearly detect
this component at high energies where it can dominate the emission.

\acknowledgments
We acknowledge useful comments by the referee on an earlier version
of the paper. This work was supported by NASA Grant NAS5-4623 at UCSD.
Archival data analysis reported in this paper was aided by tools
provided by NASA/HEASARC.

\section{References}

\def\ref{\par\noindent\hangindent 20pt}
\vglue 0.5truecm

\ref{Baumert, L.D. 1971, Lecture Notes in Mathematics, v. 182}

\ref{Belloni, T, 1998, New Astr. Reviews, 42, 585}

\ref{Belloni, T., and Hasinger, G., 1990, A\&A, 230, 103}

\ref{Bhattacharya, D. \ea 1994, ApJ, 437, 173}

\ref{Boller, Th. \ea 1992, A\&A 261, 57}

\ref{Bookbinder, J., Cowie, L.L., Krolik, J.H. \ea 1980, ApJ 237, 647}

\ref{Butcher, J.A. \ea 1997, MN 291 437B}

\ref{Cappi, M. \ea 1999, A\&A, 350, 777}

\ref{Dahlem, M., Weaver, K.A., \& Heckman, T.M. 1998, ApJS, 118, 401}

\ref{Dahlem, M. \ea 2000, ApJ, 538, 555}

\ref{Deeter, J.E. 1984, ApJ, 261, 337}

\ref{Edelson, R., \& Nandra, K. 1999, ApJ 514, 682}

\ref{Fabbiano, G. 1988a, ApJ 330, 672}

\ref{Fabbiano, G. 1988b, ApJ 324, 749}

\ref{Goldshmidt, O., Rephaeli, Y. 1995, ApJ, 444, 113}

\ref{Gruber, D.E. \ea 1996, A\&AS, 120, 641}

\ref{Gruber, D.E. \ea 2001, ApJ, 562, 499}

\ref{Immler, S. \& Wang, Q.D., 2001, ApJ, 554, 202}

\ref{Iyomoto, N., \& Makishima, K. 1999, Astron. Nachr., 320, 300I}

\ref{Jahoda, K., \ea 1996, Proc. SPIE, 2828, 59}

\ref{Kaaret, P., \ea 2001, MN, 321, L29}

\ref{Markowitz, A., \& Edelson, R. 2001, ApJ, 547, 684}

\ref{Matsumoto, H., \& Tsuru, T.G. 1999, PASJ, 51, 321}

\ref{Matsumoto, H., \ea 2001, ApJ, 547, L25}

\ref{Moran, E.C., \& Lehnert, M.D. 1997, ApJ 478, 172}

\ref{Nolan, P.L., \ea 1981, ApJ 246, 494)}

\ref{Ohashi, T. et al. 1990, ApJ 365, 180}

\ref{Pellegrini, S. \ea 2000, A\&A, 353, 447}

\ref{Persic, M., \& 1998, A\&A 339, L33}

\ref{Persic, M., \& Rephaeli, Y. 2002, A\&A, 382, 843}

\ref{Pietsch, W., \ea 2001, A\&A, 365, L174}

\ref{Ptak, A., \& Griffiths, R. 1999, ApJ 517, L85}

\ref{Ptak, A., Serlemitsos, P., Yaqoob, T. \ea 1997, AJ, 113, 1286}

\ref{Rephaeli, Y., Gruber, D., Persic, M. \ea 1991, ApJL 380, L59}

\ref{Rephaeli, Y., Gruber, D., Persic, M. 1995, A \& A, 300, 91}

\ref{Risaliti, G., Maiolino, R., \& Salvati, M. 1999, ApJ, 522, 157}

\ref{Rothschild, R.E., \ea 1998, Space. Sci. Instr., 4, 265}

\ref{Schaaf, R., \ea 1989, A\&A, 336, 722}

\ref{Shafer, R. 1983, PhD Dissertation, University of Maryland}

\ref{Strickland, D.K., \ea 2000, AJ, 120, 2965}

\ref{Tsuru, T. et al. 1990, Publ. Astron. Soc. Japan, 42, L75}

\ref{Weaver, K.A., Heckman, T.M. \& Dahlem, M. 2000, ApJ 534, 684}

\ref{White, N.E., Swank, J.H. \& Holt, S.S. 1983, ApJ 270, 711}
\end{document}